\documentclass[sigconf]{acmart}

\usepackage[utf8]{inputenc}

\usepackage{subfigure}
\usepackage{graphicx}
\usepackage{caption}
\usepackage{tabularx}
\usepackage{multirow}

\AtBeginDocument{%
  \providecommand\BibTeX{{%
    \normalfont B\kern-0.5em{\scshape i\kern-0.25em b}\kern-0.8em\TeX}}}

\setcopyright{acmcopyright}
\copyrightyear{2019} 
\acmConference[WEIHC '19]{Workshop on HCI Education - Brazilian Symposium on Human Factors in Computing Systems}{October 21--25, 2019}{Vit\'oria - ES, Brazil}
\acmBooktitle{Extended Proceedings of the 18th Brazilian Symposium on Human Factors in Computing Systems (IHC '19), October 21--25, 2019, Vit\'oria - ES, Brazil}
\acmPrice{15.00}
\acmDOI{10.1145/3357155.3358441}
\acmISBN{978-1-4503-6971-8/19/10}

\begin{document}

\title[HCI Support Card]{HCI Support Card: Creating and Using a Support Card for Education in Human-Computer Interaction}

%
\author{Lesandro Ponciano}
\affiliation{%
  \institution{Pontifical Catholic University of Minas Gerais}
  \streetaddress{Av. Dom José Gaspar, 500, Coração Eucarístico}
  \city{Belo Horizonte}
  \state{Minas Gerais, Brazil}
  \postcode{30535-901}}
\email{lesandrop@pucminas.br}

\renewcommand{\shortauthors}{Lesandro Ponciano}

\begin{abstract}
Support cards summarise a set of core information about a subject. The periodic table of chemical elements and the mathematical tables are well-known examples of support cards for didactic purposes. Technology professionals also use support cards for recalling information such as syntactic details of programming languages or harmonic colour palettes for designing user interfaces. While support cards have proved useful in many contexts, little is known about its didactic use in the Human-Computer Interaction (HCI) field. To fill this gap, this study proposes and evaluates a process for creating and using an HCI support card. The process considers the interdisciplinary nature of the field, covering the syllabus, curriculum, textbooks, and students' perception about HCI topics. The evaluation is based on case studies of creating and using a card during a semester in two undergraduate courses: Software Engineering and Information Systems. Results show that a support card can help students in following the lessons, remembering and integrating the different topics studied in the classroom. The card guides the students in building their cognitive maps, mind maps, and concept maps to study human-computer interaction. It fosters students' curiosity and permanent engagement with the HCI topics. The card usefulness goes beyond the HCI classroom, being also used by students in their professional activities and other academic disciplines, fostering an interdisciplinary application of HCI topics.
\end{abstract}


\maketitle

\section{Introduction} 

\textit{Support cards} are guides that organise a set of information about a subject in a way that facilitates and speeds up the recall of a given topic. They are widely used as an educational resource~\cite{Cassebaum:1971,anderson1995creative,cone2003benefits,balkin2017pocket}. In basic education, the ``Math reference card'' is an example of support card that organises the arithmetic operations of addition, subtraction, multiplication, and division~\cite{anderson1995creative}. Support cards are used for education in many areas, such as paediatric education~\cite{balkin2017pocket}, physics education~\cite{cone2003benefits}, and chemistry education~\cite{Cassebaum:1971}. Professionals working with technology also use support cards in their daily activities. For example, support cards are used to recall activities of a software development process~\cite{james2010scrum}, remember syntactic details of a given programming language~\cite{Short2004}, and consult colour palettes used in the design of user interfaces~\cite{stein2000web}. Therefore, the multidisciplinary and practical use of support card is well reported in the literature. 

While support cards have proved useful in many contexts, little is known about its use and its effectiveness in the teaching-learning process in the area of Human-Computer Interaction (HCI). This area integrates an interdisciplinary body of theories, recommendations for design (guidelines), heuristics, and evaluation methods that must be remembered or constantly consulted during the learning process, and during some systems' design and evaluation activities. This kind of information must be known and understood by the students, but not necessarily memorised. Inspired by their use in other disciplines, this study investigates the role that support cards can play in this context.

This study discusses the use of support cards for didactic purposes in HCI teaching and learning. A process for creating and using a support card in the HCI discipline is proposed. The process considers the interdisciplinary nature of HCI, covering the syllabus, curriculum, textbooks and students’ perception about HCI topics. The evaluation is based on case studies of creating and using a card during a semester in HCI classes that are parts of two undergraduate courses: Software Engineering and Information Systems. In doing so, three research questions are answered: (1) how to draw a support card to be used as a didactic resource in HCI classroom? (2) what utilities do students perceive in using the card? (3) how useful is the card beyond the HCI classroom?

Results from observations and questionnaires answered by 37 students show that the card can be an important didactic resource to HCI students. The card helps students in following the classes and integrating the topics of the discipline. It adds to many other pedagogical resources used in classroom~\cite{Lima:2019,Ponciano:weihc:2018}. The card also has interdisciplinary application. Students used the card in other classes that involve the design and evaluation of interactive systems, such as Software Development Laboratory, Software Testing, and Completion of Course Work. The card was also useful in professional activities, such as Supervised Internship.

\section{Background and Related Work} 

In this section, we analyse the relevant literature and related work about support cards and HCI education. 

\subsection{The Multidisciplinary Use of Support Cards} 

Support cards summarise a set of core information about a subject and are a quick way of accessing information about a subject or task. The term ``support card'' is sometimes used as a synonym for ``quick reference card'' and ``pocket card'' terms. All these terms refer to an instrument that can be easily handled, and that allows the person to find a topic quickly. Another commonly used term is ``cheat sheet'', which is a kind of support card often made by the student for unauthorised consultation during an exam.

In physics education, students are encouraged to create their own support cards, putting down anything they think will be useful, such as notes, formulas, and constants~\cite{cone2003benefits}. Differently, the card addressed in this study is a didactic resource made by a teacher to support students during their study and practices of interactive systems' design and evaluation. This meaning and this use of support card are similar in other disciplines, such as the ``Math reference card''~\cite{anderson1995creative} used in basic education, and the ``Periodic Table of Chemical Elements''\footnote{In chemistry education, the ``Periodic Table of Chemical Elements'' is a support card with visual systematic arrangement of the chemical elements ordered by atomic number, electron configuration, and recurring chemical properties.}~\cite{Cassebaum:1971} used in chemistry education. 

In some disciplines, the support card can also be used in exams and activities in the classroom~\cite{cone2003benefits}. Thus, students did not worry about memorising everything; they just check their cards. This form of work is much closer to professional situations where students must solve problems, but they usually have reference material available for consultation. Support card utilities can go beyond the process of absorbing knowledge to include hands-on activities. For example, a portable support card can improve paediatric resident education in comprehensive care for children nearing the end of life~\cite{balkin2017pocket}. The card may be a convenient, simple, and useful instrument to be used in these practical contexts. 

\subsection{HCI Education} 

Students' competence in Human-Computer Interaction involves the classic model of knowledge, skills and attitudes (KSA)~\cite{hewett1992acm,BAARTMAN2011125}. \textit{Knowledge} is the condition of being aware of something, retaining and processing information. To gain knowledge about HCI, students must learn about concepts, methods, and theories that guide the creation of strategies, guidelines, and recommendations for design. \textit{Skill}, in turn, is how to do something, performing activities and tasks on time and precisely. Applying heuristics and guidelines in designing a specific system is an example of HCI skill. Finally, \textit{attitude} is forming a new or different viewpoint or belief about a subject. Changing and adapting HCI guidelines for different contexts are attitudes.

Gaining knowledge, skill and attitudes in the HCI area is challenging for many students~\cite{Wilcox:2019,dym2005engineering}. Studies have been conducted to find strategies to make this learning process easier. Three types of strategies focused on skills can be highlighted~\cite{mccrickard2004design,Wilcox:2019}: (1) leading students to review and discuss stories describing project situations ({\it history reviews}); (2) conducting controlled studies in which students are led to solving practical problems ({\it problem solving cases}); (3) engaging students in situations where they must make decisions and justify their choices ({\it decision-making cases}). Students also must have attitude, thinking critically about the design, development, and evaluation of interactive systems~\cite{mccrickard2004design, churchill2013teaching}. Strategies for achieving this end in the classroom include, for example: structured debates about human values ~\cite{Ponciano:weihc:2018}; lessons that integrates theories, drawings and music with active participation of the students~\cite{Silva:weihc:2018}; and the institutionalisation of interdisciplinary activities~\cite{Britto:weihc:2018}. 

These strategies presuppose that students have at least a basic knowledge of the key topics in the area. Knowledge is a building block for ability and attitude. Gaining knowledge is synonymous to retain information, usually through lectures and reading textbooks. For the new generations of students, who have been born with available information and communication technologies~\cite{Nelson:2017,Erdogmus:2017,Boscarioli:2016}, memorising information may not be necessary, especially if it is readily available for consultation. In this paper, we propose the use of support card as a consultation didactic instrument to help students in gaining knowledge about the core HCI topics.

\section{Process of Creating and Using a Support Card}

This section presents a process for creating and using a support card for HCI education. The process focuses on three main questions that one must answer in creating and using a card: \textit{What are the main requirements of the support card? How to define the topics to be covered on the card? How to organise the topics on the card?} and \textit{How to use the card in the classroom?} In the following paragraphs, we discuss our approach to address these questions.

We set out four main requirements for an HCI support card: (1) the card must be comprehensive of the content of the syllabus; (2) the card must preserve the interdisciplinary nature of the area; (3) the card must be useful to students in the classroom and homework; and, (4) the card should be easy to handle and read. The card must be comprehensive so that it can be used throughout the lessons and allows the student to connect the studied topics. The card should not be only technical; instead, it should highlight knowledge coming from other areas, preserving and contextualising interdisciplinary topics. Finally, to be used effectively, the card must provide value to students, such as serving as a quick reference to core HCI topics, an aid to the recall of some concept/method, and a study guide for exams. To be easy to read and handle, the card can be provided to students as images that can be viewed in smartphone/tablet or printed on a two-sided paper laminated with plastic material on both sides.

It is impossible to cover all syllabus topics on a one-page card. Some topic prioritisation must be done. At a macro level, the card should cover the major topics defined in the discipline's syllabus, textbooks, and curriculum. At a micro level, the card should cover topics that are harder to remember, and topics that serve as anchors for other topics. The card works as a support instrument for student's memory, and topics included on the card serve as triggers for other topics that are not included on the card. So, interdisciplinary topics are especially important to be covered on the card. 

\begin{figure}[htb] 
\centering 
\includegraphics[scale=0.143]{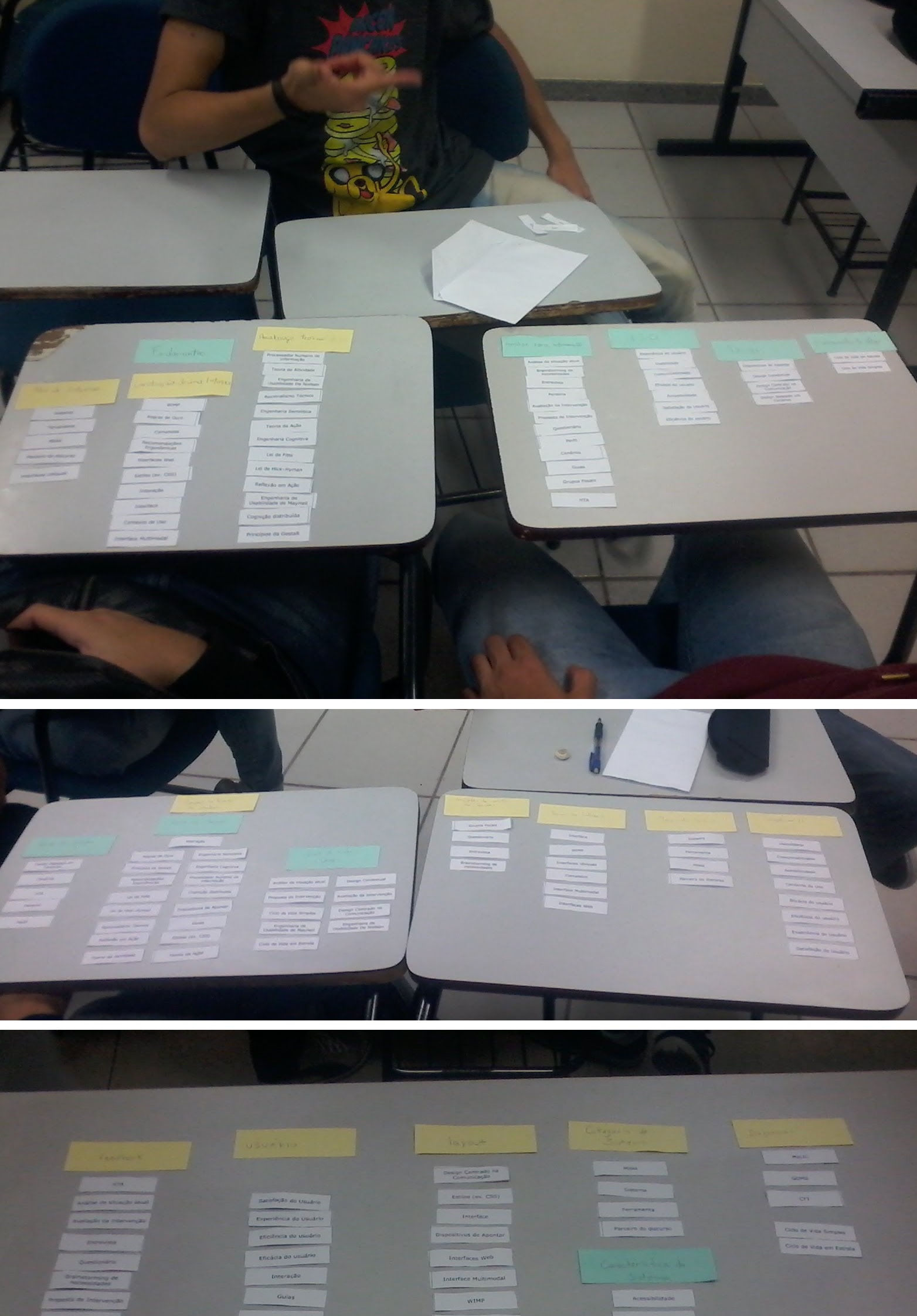} 
\caption{Card sorting sessions. Students are asked to cluster and categorise HCI topics in way that makes sense to them.} 
\Description{Card sorting sections} 
\label{fig1} 
\end{figure}

The organisation of the topics on the card must be intuitive for students so that they can use the card effectively. This can be done by using the ``card sorting technique''. This technique is known in the HCI area since it is used in designing systems' information architecture~\cite{Wood:2008}. As part of creating a support card, this technique can be used to ask students to organise topics of the discipline into groups that make sense to them, and label those groups, forming categories of HCI topics (as exemplified in Figure~\ref{fig1}). The card sorting outcomes inform how students would cluster and label HCI topics. Of course, it must be done when students have already studied the topics. Results from the card sorting dynamics are insights for the information architecture of the next versions of the support card, which will be used by future students of the discipline. Thus, the card is renewed with each HCI course.

The support card is made available to students after a few classes, when some topics have already been taught to them. For example, the card can be released after the first 8 classes in an HCI course consisting of 68 classes. The idea is that, based on what has already been studied, students can see how the studied topics are summarised on the card. The card should be used in the classroom. An association between the subject of the class and the topics on the card should be made. For example, the card may contain a topic (represented by a word or picture) associated with each set of slides or each book chapter. Teachers can highlight where each subject of the class appears on the card. In hand-on activities, such as Nielsen's Heuristic Evaluation of a system~\cite{Nielsen:1990}, students can consult the support card to remember the heuristics. Teachers can also stimulate a creative use of the card, such as asking students to tell a story by articulating HCI concepts shown on the card.

\section{Evaluation} 

Following the approach described in the previous section, we created a support card and used it on two courses. In this section, we first present the card. Then, we discuss the method employed to evaluate the card and its use. Finally, we discuss the results. 

\subsection{The Support Card} 

\begin{figure*}[htb] 
    \centering 
    \subfigure[Face \textit{A}, predominance of conceptual and theoretical HCI topics]{ 
        \includegraphics[scale=0.14]{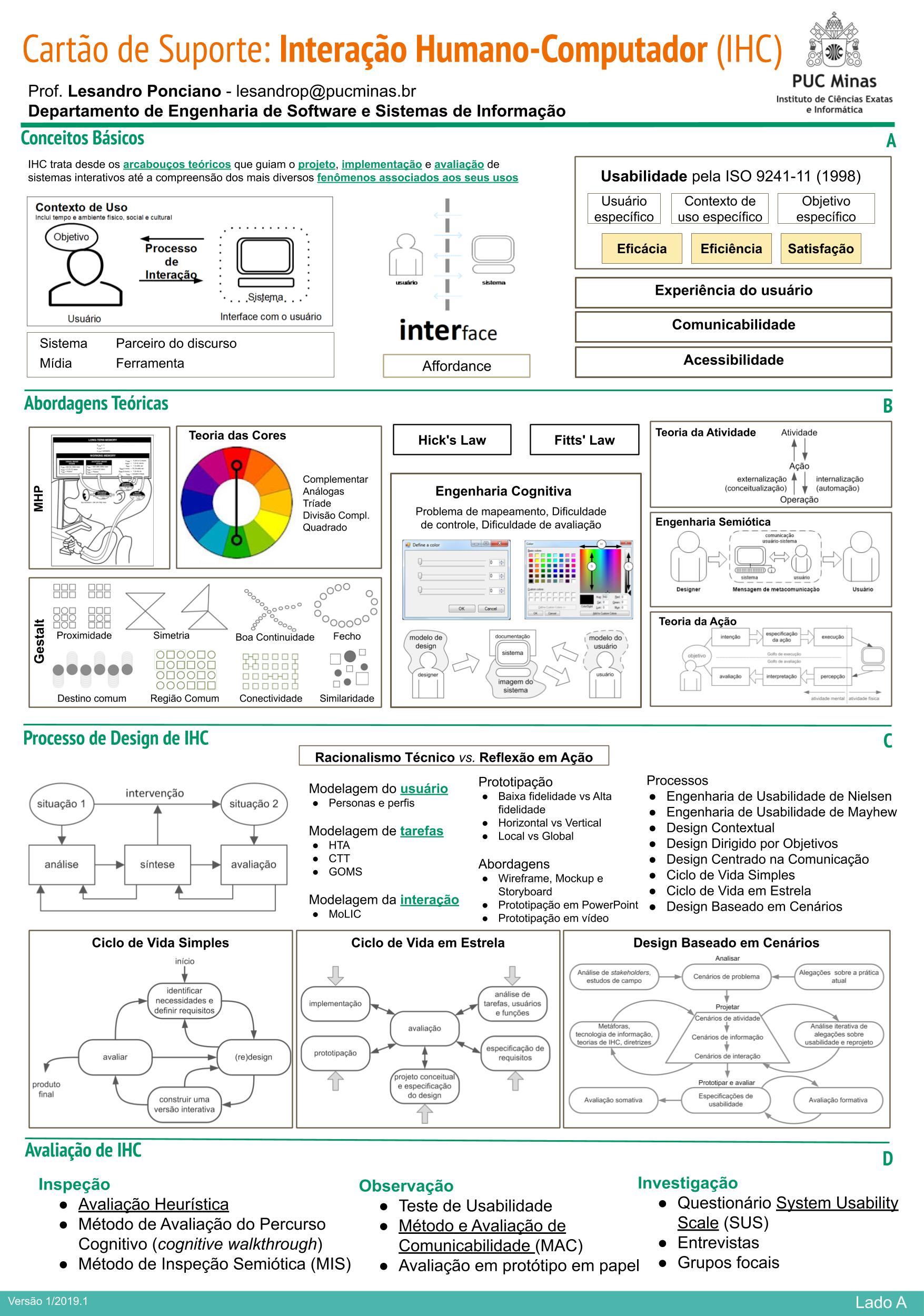} 
    } 
    \subfigure[Face \textit{B}, predominance of practical HCI topics]{
        \includegraphics[scale=0.14]{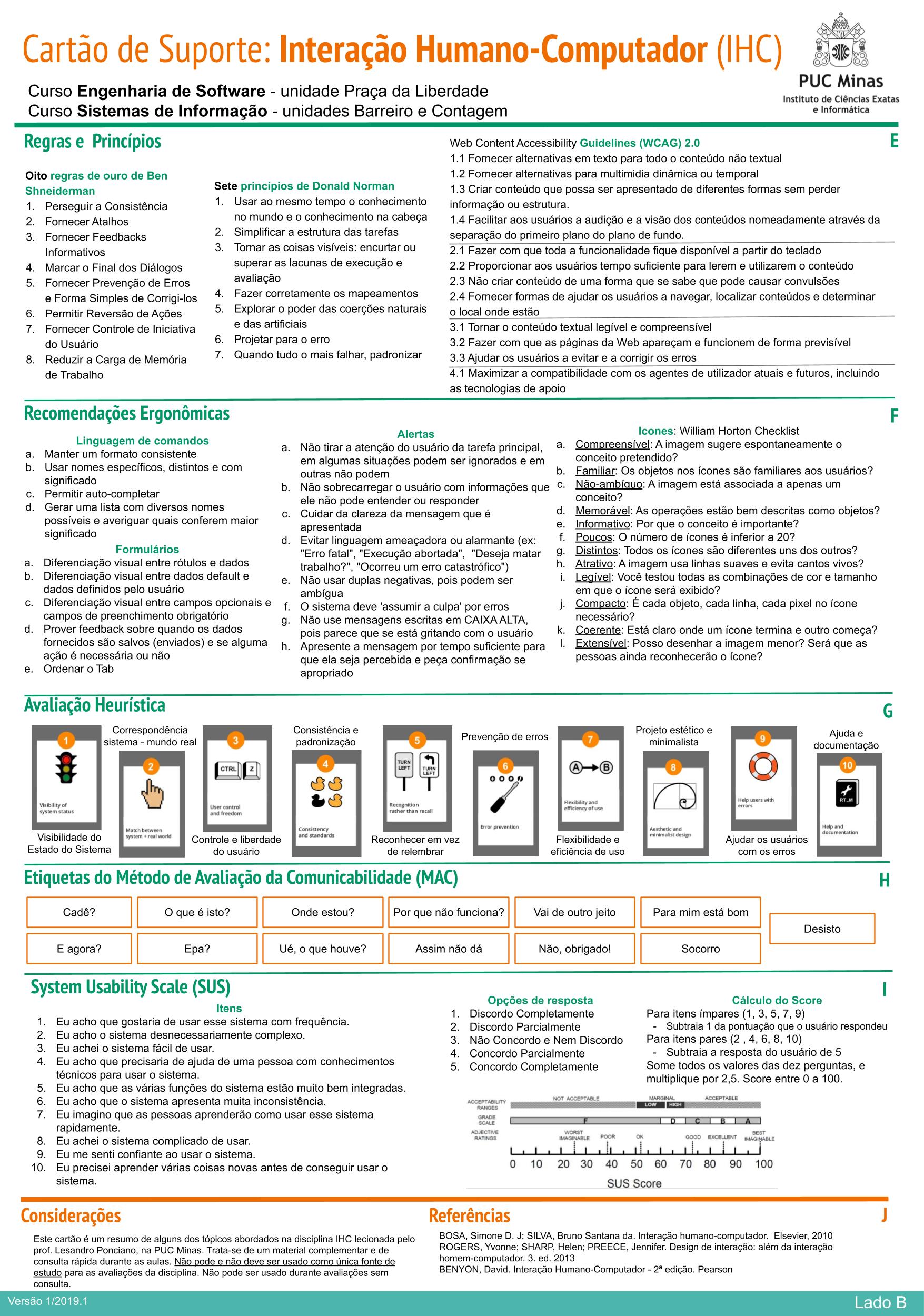} 
    } 
    \caption{Overview of the layout of the HCI Support Card. Full size and high-resolution image can be download at \url{www.lsd.ufcg.edu.br/~lesandrop/cartaoIHC}.} 
    \label{fig2} 
\end{figure*}

Figure~\ref{fig2} shows an overview of card's layout. The card has two faces: face \textit{A}, and face \textit{B}. Topics on face \textit{A} are more conceptual and theoretical than those on face \textit{B}. Face \textit{A} brings together HCI concepts and theories. Face \textit{B} brings together topics directly used when designing or evaluating interactive systems, such as: ergonomic guidelines, golden rules, heuristics, and guidelines for icons.

The card covers four major groups of topics: \textit{Basic Concepts}, including interface, affordance, usability, and communicability; \textit{Theoretical Approaches}, including Activity Theory, Action Theory, Colour Theory, and Semiotic Engineering; \textit{Design Process}, including Usability Engineering, and Scenario-based Design; and, \textit{Evaluation Methods}, including Heuristic Evaluation, System Usability Scale, and Semiotic Inspection.

The material presented on the card is based on the textbooks by Barbosa e Silva~\cite{barbosa2010interacao}, and by Rogers, Sharp and Preece~\cite{rogers2011interaction}, which are the main textbooks used in the courses in which the card was made. We also considered topics indicated in the ACM SIGCHI Curricula for Human-Computer Interaction~\cite{hewett1992acm}.

\subsection{Research Methods} 

This study follows a case study approach~\cite{runeson2012case}. The case studies were conducted in Software Engineering and Information Systems undergraduate courses, which are taught in two campus of the Pontifical Catholic University of Minas Gerais (PUC Minas). The card was created at the end of the second half of 2018 by using insights from the card sorting dynamics. The card was used in classes during the first semester of 2019 in the Information Systems and Software Engineering undergraduate courses. It was made available to the students as images in the Portable Network Graphics (PNG) format, so students can use it printed or digital. The card is fully readable when printed in A4 size. In digital format, the image size can be enlarged or reduced according to student's preferences.

Software Engineering and Information Systems undergraduate courses are composed of 8 periods, each period being equivalent to one semester. HCI discipline is currently located in the fifth period in Information System course, and in the fourth period, in Software Engineering course. In both courses, HCI discipline has the same syllabus, has in total 68 classes each of 50 minutes, and has Requirements Engineering discipline as a prerequisite.

The evaluation was made through teacher's class notes and a questionnaire. Questions of the questionnaire are shown in Table~\ref{tab:questions}. The questionnaire was answered by students after using the card in classroom for 4 months. Students' participation was informed, voluntary, and anonymous. Altogether, $37$ students answered the questionnaire, being $17$ students from the Software Engineering course (hereafter they are identified with codes ranging from SE-P1 to SE-P17) and $20$ students from the Information Systems course (hereafter identified with codes ranging from IS-P1 to IS-P20).

\begin{table*}[htb] 
\caption{Questions to assessment students' perception about the use of support card in the HCI classroom.} 
\label{tab:questions}
\begin{tabularx}{\linewidth}{lXX} 
\toprule 
\multicolumn{1}{c}{Item of evaluation}&\multicolumn{1}{c}{Question}&\multicolumn{1}{c}{Answer format/options}\\ 
\midrule 
\multirow{3}{*}{Engagement with the card} & Do you estimate that you consulted the card how many times throughout the HCI discipline?& (a) None; (b) 1 to 5 times; (c) 6 to 10 times; (d) 10 to 15 times; (e) More than 15 times\\ \\ 

& In what format did you use the card? & (a) digital format (e.g., smartphone, tablet, computer); (b) physical format (e.g., printed on paper)\\ \midrule 

\multirow{3}{*}{Usefulness in the HCI discipline}& Would you recommend the card to other HCI students?& Yes/No answer \\ \\ 

& Do you think the card should continue to be made available to students and used in the next classes of the HCI discipline? & Yes/No answer, with an open-ended justification\\ \\ 

& How much do you agree with the statement ``The card summarises the main topics studied in the HCI discipline''& Five-point Likert scale answer~\cite{likert1932technique} \\ \midrule 

\multirow{3}{*}{Usefulness beyond HCI discipline}& Was the card useful for you in any professional activity? (such as in the company in which you work, on the internship, or something equivalent) & Yes/No answer, with an open-ended justification\\ \\ 

& Has the card been helpful to you in any discipline other than the HCI discipline?&Yes/No answer, with an open-ended justification\\ \\ 

& Do you think the card will be of any use to you after the completion of the HCI discipline? & Yes/No answer, with an open-ended justification\\ \midrule 

\multirow{3}{*}{Suggestion of improvement}& Is there anything about HCI that you do not have on the card and that you believe should be added to the card? If yes, give us an example. & Open-ended response\\ \\ 

& Do you have anything else to tell us about the card or about its use in the classroom? If yes, please, write below. & Open-ended response\\  
\bottomrule 
\end{tabularx} 
\end{table*}

\subsection{Results} 

{\bf The usefulness of the support card in the HCI education.} Students have used and recommended the card. A total of $12$ ($33\%$) students used the card in physical format (e.g. printed on paper), $20$ ($53\%$) students used the card in digital format, and $5$ ($14\%$) students used it in both formats. A total of 11 (30\%) students used the card $10$ or more times throughout the semester. All the $37$ students ($100\%$) answered that would recommend the card to other HCI students.

Altogether, a total of $36$ students ($97\%$) suggested that the support card should continue to be used in the HCI discipline. Of these students, a total of $29$ ($78\%$) students provided some justification for their suggestion. Some of such justifications are detailed below. 

\begin{itemize} 
\item Students explain that the support card allows for a quick search and exploration of the topics of the discipline. For example, students declare that: \textit{``The card helps a lot to absorb the contents of the discipline, besides allowing consulting the content quickly''} (IS-P3), \textit{``It assists in quick consultation of rules, evaluation methods, etc. in the process of evaluation and construction of the system, serving a bit like a mental map for the exam''} (IS-P15), \textit{``The card serves as a basic guide to remember the topics''} (SE-P7); \textit{``It helps me in connecting the idea to the name; sometimes I remember the content, but I forget the name, the card helps a lot.''} (SE-P16). 

\item Students explain that the support card makes it easy to study and organise the topics of the discipline. For example, students declare that: \textit{``The card helps at the time of the studies making it easy to recall the contents''} (IS-P14); \textit{``It allows us to remember the content and even to better summarise the contents at the time of studying, highlighting the main subjects within HCI discipline''} (IS-P9); \textit{ ``It is useful to organise content in my mind''} (SE-P1); and \textit{``The card is great for disciplines like HCI - full of details, rules and keywords''} (SE-P6). 
\end{itemize} 

From the teacher's perspective, one effect of using the support card in classroom is that it arouses \textit{curiosity} and allows for a \textit{permanent engagement} with HCI topics. At the beginning of the semester, when the card is made available, students explore the entire card and are curious about the topics on it ({\it what is this? when will we study this?}). During the semester, when handling the card in the classroom, the student sees, reads, thinks about small elements of the whole discipline. This constant contact causes a permanent engagement with the HCI topics. 

{\bf The usefulness of the support card beyond the HCI classroom (in other undergraduate disciplines or professional activities).} Students indicate that they used the card in professional activities. For example, the IS-P9 student says ``\textit{When I am developing a user interface, I try to remember the concepts of HCI, such as gestalt principles, and the card helps me to remember}''. Other justifications are associated with knowledge gained in the classes, for example: ``\textit{I showed some errors when developing the ERP [Enterprise Resource Planning] where I work}'' (IS-P12);  ``\textit{In a development team, at the company, we discuss about aesthetic scenarios}'' (IS-P14); and ``\textit{[I consider the topics] in the use of colours for an application}'' (IS-P11),

Students also used the card to remember and apply HCI topics in other disciplines of the course, such as the following disciplines: \textit{Interdisciplinary Software Work IV} (SE-P3 and SE-P6); \textit{Software Development Laboratory} (SE-P6); \textit{Completion of Course Work} (SE-P10 and SE-P16); and, \textit{Software Testing} (IS-P7). The student IS-P7 explains ``\textit{the card is useful in Software Testing because it has concepts that help in understanding the point of view of a user}''. The card summarises some evaluation methods such as Nielsen's Heuristic Evaluation and System Usability Scale (SUS), making it easier to associate HCI topics and software testing in general. The use of the card by students in other disciplines can go beyond the semester in which they are studying HCI: \textit{``At the end of the course, the slide is rarely consulted, but the card can be''} (IS-P8).

{\bf Suggestions of improvements on the card.} Two students provided suggestions about the layout of the card: \textit{``I think a point to change would be the subjects being in order of teaching in the classroom''} (SE-P3), and \textit{``Should have a pocket version''} (IS-P12). This type of suggestion is important to understand how the student would like the card to be, how to improve the card in future versions, and how to avoid misunderstanding about card's purpose.

{\bf Limitations.} Our proposal and results have limitations that should be highlighted. Although the card has been used in classes of two different courses (Information Systems and Software Engineering), it is unknown if students' perception would be similar in HCI disciplines taught as part of other courses, such as Computer Engineering, Art, Digital Games Development, and Computer Science. As the support card is an auxiliary resource and of optional use, it can be used only by those students who find its use beneficial.

\section{Discussions and Conclusions}

Support card is a pedagogical resource widely used in many disciplines. In this work, we report the experience of creating and using a support card in HCI classes in Information Systems and Software Engineering courses. We discuss the main requirements of the support card, how to define the topics to be covered on the card, how to organise the topics on the card, and how to use the card in the classroom.

Our results show the importance of support card as a didactic resource in HCI education. It helps the students (1) to follow the lessons, (2) to integrate the different topics of the course, and (3) to make it easy an interdisciplinary application of studied topics. The card works as one of the inputs for students to build their cognitive maps, mind maps, and concept maps in studying human-computer interaction. It stimulates students' curiosity and engagement with HCI topics. The use of the card goes beyond the HCI discipline; it is also used by students in their professional activities and other academic disciplines.

One cannot expect a single and definitive support card to the HCI field. HCI is a dynamic field, so the card should be adapted continually following such dynamic and the needs of each course. Thus, the card is a dynamic instrument, changing with the context of the course in which it is used. The process described in this paper can be followed to create and adapt a support card to each context. To be effective, the support card should also be a ``student-oriented'' card, being continually adapted to students' needs. Feedback from student should be continually obtained and considered so that the card remains effective overtime. Finally, the card is complementary; it adds to many other pedagogical resources used in HCI education.

We suggest the following questions to be investigated in future work. What is the most effective method for establishing which syllabus topics should be highlighted on the card? Can interaction features make even more engaging the digital use of the card? It is also relevant to create a card totally focused on the practical work of designing and evaluating interactive systems. Doing so will require a less pedagogical and more market-oriented approach. 

Future work can also investigate the use of support card beyond the HCI teaching-learning process. A theoretical and general understanding of the use of this type of artefact is desired. It may be useful in many situations where a human being performs tasks that require remembering or processing an amount of information that goes beyond the limits of the cognitive system~\cite{simon1972theories}, such as some human intelligence or human computation tasks~\cite{ponciano2014considering}. We hope this work will lead to new studies in this direction.

\bibliographystyle{ACM-Reference-Format}
\bibliography{sample-base}
\end{document}